\documentclass[12pt]{iopart}
\usepackage{iopams}  
\usepackage{graphicx}
\usepackage{ulem}

\begin{document}

\title[Relativistic calculation of K$\beta$ hypersatellite
transitions]{Relativistic Calculation Of K$\beta$ Hypersatellite
  Energies and Transition Probabilities for Selected Atoms with $13$
  $\leq$ $Z$ $\leq80$}

\author{A. M. Costa\dag\S 
\footnote[8]{To whom correspondence should be addressed
  (amcosta@fc.ul.pt) } ,\ 
M. C. Martins\dag\S, \ 
J. P. Santos\ddag\S,\ 
P. Indelicato$\|$ and 
F. Parente\ddag\S}

\address{\dag\ Departamento de F{\'\i}sica, Faculdade de Ci{\^e}ncias,
  Universidade de Lisboa, Campo Grande, 1749-016 Lisboa, Portugal}

\address{\ddag\ Departamento de F{\'\i}sica, Faculdade de Ci{\^e}ncias e
  Tecnologia, Universidade Nova de Lisboa, Monte de Caparica, 2825-114
  Caparica, Portugal}

\address{\S\ Centro de F{\'\i}sica At{\'o}mica da Universidade de
  Lisboa, Av. Prof. Gama Pinto 2, 1649-003 Lisboa, Portugal}

\address{$\|$\ Laboratoire Kastler Brossel,
 \'Ecole Normale Sup\' erieure; CNRS;  Universit\' e P. et M. Curie -
 Paris 6,  Case 74; 4, place Jussieu, 75252 Paris CEDEX 05, France}
 

\begin{abstract}
  Energies and transition probabilities of K$\beta$ hypersatellite
  lines are computed using the Dirac-Fock model for several values of
  $Z$ throughout the periodic table. The influence of the Breit
  interaction on the energy shifts from the corresponding diagram
  lines and on the K$\beta_{1}^{\rm h}$/K$\beta_{3}^{\rm h}$ intensity
  ratio is evaluated. The widths of the double-K hole levels are
  calculated for Al and Sc. The results are compared to experiment and
  to other theoretical calculations.

\end{abstract}
 



\vspace{1cm}
\noindent {\bf \today }

\maketitle

  
\section{Introduction} 
 
Energies and transition probabilities of K$\beta$ hypersatellite lines
are evaluated in this work for selected atoms throughout the periodic
table. In a previous paper ~\cite{1105}, we reported on calculated
values of K$\alpha$ hypersatellite line energies for atoms with
$12\leq Z\leq30$.
 
A hypersatellite line is an X-ray line for which the initial state has
two vacancies in the same shell. This is the case, for example, when a
double ionized K-shell state decays through the transition of one
M-shell electron.  Lines corresponding to 1s$^{-2}$ $\to$
1s$^{-1}$3p$^{-1}$ transitions, where $n \ell$$^{-k}$ means $k$
vacancies in the $n \ell$ subshell, are called K$\beta_{1,3}^{\rm h}$
hypersatellite lines.
 
Atoms where a whole inner shell is empty while the outer shells are
occupied were first named hollow atoms by Briand \etal ~\cite{888} and
are of great importance for studies of ultrafast dynamics in atoms far
from equilibrium and with possible wide-ranging applications in
physics, chemistry, biology, and materials science ~\cite{889}.
 
Very scarce experimental data exist on energies of K hypersatellite
lines, due to the low probability of creation of the initial two K
holes.  Briand  \etal ~\cite{911} used a coincidence method to
study the K hypersatellite spectrum in Ga and measured a 390 eV energy
shift for the K$\beta^{\rm h}$ lines relative to the corresponding
diagram lines. Energies of K$\beta^{\rm h}$ hypersatellite lines were
later measured by Briand \etal~\cite{900,899} for Mn, Ga and
In. Diamant \etal ~\cite{1015} obtained high resolution
K$\beta_{1,3}^{\rm h}$ and K$\alpha^{\rm h}$ hypersatellite spectra of
Fe, using monochromatized synchrotron radiation photoexcitation.
Similar work has been performed for Cr K$\beta^{\rm{h}}$
hypersatellite spectra ~\cite{rep2000}. More recently, this work has
been extended to Ti ~\cite{1208}.
 
On the theoretical side, calculations of the energies and transition
probabilities of K$\beta_{1,3}^{\rm h}$ hypersatellite lines have been
performed \cite{308} using a Dirac-Hartree-Slater
approach. This approach employs a local approximation to the atomic
potential. With the wave functions obtained from this potential,
perturbation calculations were then used to obtain the energies.
Diamant \etal ~\cite{1015} performed relativistic multi-configuration
Dirac-Fock calculations for the K$\beta_{1,3}^{\rm h}$ hypersatellite
lines of Fe to compare with their own experimental findings.
 
The energy shifts between the K$\beta_{1,3}^{\rm h}$ hypersatellite
lines and the corresponding K$\beta_{1,3}$ diagram lines (as the
K$\alpha^{\rm h}$-K$\alpha$ ones) are considered to be very sensitive
to the Breit interaction.  Chen \etal ~\cite{308} calculated the
influence of the Breit interaction on these shifts, using a
mathematical form appropriate for the local approximation employed.
To check the importance of the Breit interaction, values of these
shifts as well of the K$\beta_{1}^{\rm h}$/K$\beta_{3}^{\rm h}$
intensity ratios were calculated in this work for selected values of
$Z$, with and without inclusion of the this interaction.

Using the results of our calculations, we were able to obtain the
widths of the $1s^{-2}$ two-hole levels of Al and Sc.
 

\section{Calculation of atomic wave functions and energies} 
 
Bound state wave functions and radiative transition probabilities were
calculated using the multi-configuration Dirac-Fock program of J. P.
Desclaux and P. Indelicato ~\cite{32,62,62a}. The program was used in
single-configuration mode because correlation was found to be
unimportant for the energies and probabilities transition. The wave
functions of the initial and final states were computed independently,
that is, atomic orbitals were fully relaxed in the calculation of the
wave function for each state, and non-orthogonality was taken in
account in transition probability calculations.
 
In order to obtain a correct relationship between many-body methods and 
quantum electrodynamics (QED)~\cite{47,231,229,230}, one should start from the 
no-pair Hamiltonian 
\begin{equation} 
\mathcal{H}^{\mbox{{\tiny no pair}}}=\sum_{i=1}^{N}\mathcal{H}_{D}(r_{i}%
)+\sum_{i<j}\mathcal{V}(r_{ij}), 
\label{eq:hamilnopai}
\end{equation} 
where $r_{ij}=\left\vert
  \boldsymbol{r}_{i}-\boldsymbol{r}_{j}\right\vert $, and 
$\mathcal{H}_{D}$ is the one electron Dirac operator and $\mathcal{V}$
is an operator representing the electron-electron interaction of order
one in
$\alpha$, properly set up between projection operators $\mathit{\Lambda}%
_{ij}^{++}=\mathit{\Lambda}_{i}^{+}\mathit{\Lambda}_{j}^{+}$ to avoid coupling 
positive and negative energy states 
\begin{equation} 
\mathcal{V}(r_{ij})=\mathit{\Lambda}_{ij}^{++}V(r_{ij})\mathit{\Lambda}_{ij}^{++}. 
\end{equation} 
In nonrelativistic quantum mechanics the interaction between charged
particles is described by an operator associated to the potential
energy of interaction, $V\left( r_{ij}\right)$, which is only a
function of the distance between particles, $r_{ij}$. However, in
quantum electrodynamics there exists no potential energy of
interaction that depends on the coordinates of the interacting charged
particles taken at the same time, since the charges interact not
directly, but via the electromagnetic field. The operator for the
interaction potential energy is replaced, in the first approximation,
by the scattering matrix ~\cite{12}
\begin{equation} 
\mathsf{S}_{AB\to CD}^{(2)}=e^{2}\int\overline{\psi}_{D}\left( 
y\right)  \gamma_{\mu}\psi_{B}\left(  y\right)  D^{c}\left(  y-x\right) 
\overline{\psi}_{C}\left(  y\right)  \gamma_{\mu}\psi_{A}\left(  y\right) 
\;d^{4}y\,d^{4}x, 
\label{37_4}%
\end{equation} 

After performing the time integration of (\ref{37_4}), one finds the
following well known interaction potential, in Coulomb gauge and in
atomic units,
\numparts
\begin{eqnarray}
g\left(  r_{ij}\right)   &  = \frac{1}{r_{ij}%
}\label{eq:coulop}\\ 
&  -\frac{\boldsymbol{\alpha}_{i}\cdot\boldsymbol{\alpha}_{j}}{r_{ij}%
}\label{eq:magop}\\ 
&  -\frac{\boldsymbol{\alpha}_{i}\cdot\boldsymbol{\alpha}_{j}}{r_{ij}}%
[\cos\left( \frac{ \omega_{ij}r_{ij}}{c}\right)  -1] \nonumber\\ 
&  +(\boldsymbol{\alpha}_{i}\cdot\nabla_{i})(\boldsymbol{\alpha}_{j}%
\cdot\nabla_{j})\frac{\cos\left( \frac{ \omega_{ij}r_{ij}}{c}\right)  -1}{\omega 
_{ij}^{2}r_{ij}}, \label{eq:allbreit}
\end{eqnarray}
\endnumparts
where is $\omega_{ij}$ is the energy of the photon exchanged between
the two electrons, $\boldsymbol{\alpha}_{i}$ are the Dirac matrices
and $c=1/\alpha$ is the speed of light. The first term
(\ref{eq:coulop}) represents the Coulomb interaction, the second one
(\ref{eq:magop}) is the Gaunt interaction (magnetic interaction), and
the last two terms (\ref{eq:allbreit}) stand for the retardation
operator. In this expression the $\nabla$ operators act only on
$r_{ij}$ and not on the following wave functions. By a series
expansion of the terms Eqs.~(\ref{eq:magop})-(\ref{eq:allbreit}) in
powers of $\omega_{ij}r_{ij} /c\ll1$ one obtains the Breit
interaction, which includes the leading retardation contribution of
order $1/c^{2}$. The Breit interaction is the sum of the Gaunt
interaction (\ref{eq:magop}) and of the Breit retardation
\begin{equation} 
g^{\rm{B}}\left(  r_{ij}\right)  =\frac{\boldsymbol{\alpha}_{i}\cdot 
\boldsymbol{\alpha}_{j}}{2r_{ij}}-{\frac{\left(  \boldsymbol{\alpha}_{i}%
\cdot\boldsymbol{r}_{ij}\right)  \left(  \boldsymbol{\alpha}_{j}%
\cdot\boldsymbol{r}_{ij}\right)  }{{2r_{ij}^{3}}}}. \label{eq:breit}%
\end{equation} 
In the present calculation the electron-electron interaction is described by 
the sum of the Coulomb and the Breit interaction. Higher orders in $1/c$, 
coming from the difference between Eqs.~(\ref{eq:allbreit}) and 
(\ref{eq:breit}) are treated here only as a first order perturbation. 
 
We use the Coulomb gauge as it has been demonstrated that it provides
energies free from spurious contributions at the ladder approximation
level and must be used in many-body atomic structure
calculations~\cite{846}. 
 
Finally, from a full QED treatment, one also obtains the radiative corrections 
(important for the innermost shells) to the electron-nucleus interaction 
(self-energy and vacuum polarization). The one-electron self-energy is 
evaluated using the one-electron values of Mohr and coworkers 
~\cite{115,114,116}. The self-energy screening is treated with the Welton 
method developed in Refs.~\cite{58,56,53,847}. This method yields results in 
close agreement (better than 5\%) with \textit{ab initio} methods based on QED 
~\cite{242,263,288,iam2001}, without the huge amount of effort involved. 

The vacuum polarization is evaluated as described in Ref.~\cite{914}.
The Uelhing contribution is evaluated to all orders by being included
in the self-consistent field (SCF). The Wichmann and Kroll, and
K\"{a}ll\'{e}n and Sabry contributions are included perturbatively.
These three contributions are evaluated using the numerical procedure
from Refs.~\cite{158,277}.

 
\section{Results} 
 
\subsection{Introduction} 

We calculated the energy shifts of the K$\beta_{13}^{\rm{h }}$
hypersatellite lines relative to the corresponding diagram lines for
several atoms with $13\leq Z\leq80$, and the K$\beta_{1}^{\rm{h }}$
and K$\beta_{3}^{\rm{h }}$ energy shifts, and their relative
intensities, for selected atoms with $18\leq Z\leq80$.
The wave functions of the initial and final states were computed
independently, that is, atomic orbitals were fully relaxed in the
calculation of the wave function for each state.

In what concerns the precise identification of the K$\beta_{1,3}^{\rm
  h}$ hypersatellite lines some comments are in order.

Depending on the configurations of the initial and final states, for the 
different values of $Z$, the number of transition lines that must be computed 
may range from only two, when the initial state has only closed shells, to 
several thousand, when unfilled shells exist. 

For elements with filled shells, namely Ar, Ca, Zn, Kr, Sr, Pd, Cd,
Xe, Ba and Hg the 1s$^{-2}$ ground configuration corresponds to only
one level, the $^{1}$S$_{0}$ level, and each of the K$\beta^{\rm h}$
lines is identified by a precise level transition,
  
\begin{tabular} 
[c]{ll}
K$\beta_{3}^{\rm{h}}$: & $1$s$^{-2\rm{ \ }1}$S$_{0}\to1$s$^{-1}%
3$p$^{-1}$ $^{1}$P$_{1}$\\ 
K$\beta_{1}^{\rm{h}}$: $\ $ & $1$s$^{-2\rm{ \ }1}$S$_{0}\to 
1$s$^{-1}3$p$^{-1}$ $^{3}$P$_{1}$.
\end{tabular}

By analogy with the corresponding diagram lines, the line that
corresponds to the larger transition energy value is labeled
K$\beta_{1}^{\rm{h}}$ and the other one is labeled
K$\beta_{3}^{\rm{h}}$.

For $Z\geq29$ there are two sets of transition lines, separated in
energy by more than 3 eV, corresponding to the K$\beta_{1}^{\rm{h}}$
and K$\beta_{3}^{\rm{h}}$ lines. In the first set the decay is due
mainly to the 3p$_{3/2}$ electron transition, whereas in the second
set is mainly due to the 3p$_{1/2}$ electron transition.
LS coupling dominates the level structure for elements with $Z<29$, as
can be seen in Fig.~\ref{figure} for Si. In this figure, the labels
refer to the only transitions that contribute for the spectrum.
Intercombination lines give negligible contribution.

 
\subsection{Transition probabilities}

The transition probability $W^{X}$ for the line $X$ is defined as
\begin{equation} 
  W^{X}= \frac{\sum_{i} N\left(
  i\right)W_{i}^{X}}{N\left(\gamma\right)}  
  \label{eq008}
\end{equation} 
where $\gamma$ is, in this case, a given double-K hole configuration,
$N\left( i\right) $ is a collection of excited atoms per unit volume
and $N\left( \gamma\right) = \sum_i N(i)$.
$W_{i}^{X}$ is the transition probability for the line $X$ from an
initial level $i$, defined by
\begin{equation} 
W_{i}^{X}  = \sum_{f^{X}} W_{if},
\label{eq009}
\end{equation} 
where $f^{X}$ runs over all possible final levels in the radiative
de-excitation process leading to the $X$ line, and $W_{if}$ is the
probability per unit time that an atom in the excited level $i$, will
decay to the level $f$, through the spontaneous emission of a photon.
 
For short lifetimes $\tau$ of the excited levels, compared with
characteristic creating times (the inverse of the number of
excitations per second undergone by one atom), the number of atoms
doubly ionized in the K shell created in the excited level $i$ per
unit time $C_{\gamma}\left( i\right)$ equals the rate at which the
atoms leave the level $i$, $N\left( i\right) /\tau\left( i\right)$, by
all possible transitions. Assuming that the $i$ level of a given
double-K hole $\gamma$ configuration is fed according their
statistical weight, we have
\begin{equation} 
C_{\gamma}\left(  i\right)  =C_{\gamma}\frac{g\left(  i\right)  }{g\left( 
\gamma\right)  },
\label{eq010}
\end{equation} 
where $g\left( i\right) $ and $g\left( \gamma\right) $ are the
multiplicities of the $i$ level and of the $\gamma$ double-K hole
configuration, respectively, and $C_{\gamma}$ is the number of
double-K ionised atoms created per unit time and per unit volume. From
Eq. (\ref{eq008}) and (\ref{eq010}) we obtain
\begin{equation} 
W^{X}=\frac{\sum_i g\left(  i\right)  \tau\left(  i\right) 
W_{i}^{X}  }{\sum_i g\left(  i\right)  \tau\left( 
i\right)  }. 
\end{equation}
Using 
\begin{equation} 
\tau\left(  i\right)  =\frac{1}{\sum_f
  W_{if}+ \sum_{f^{\prime}} A_{if^{\prime}}}, 
\end{equation} 
where $f$ runs for all possible final levels that can be reached by
radiative transitions, with probabilities $W_{if}$, and $f^{\prime}$
runs for all possible final levels that can be reached by
radiationless transitions, with probabilities $A_{if^{\prime}}$, we
get
\begin{equation} 
W^{X}=\frac{\sum_i \frac{g\left(  i\right)  W_{i}^{X} 
}{W_i  +A_i  }}{\sum_i \frac{g\left( 
i\right)  }{W_i  +A_i  }}.
\label{10} 
\end{equation} 
We made use of $W_i =\sum_f W_{if}$, and $A_i =\sum_{f^{\prime}}
A_{if^{\prime}}$.
 
For the elements where the two K-hole ground configuration has more
than one level, it is therefore necessary to compute, for each of
those levels, not only radiative transition probabilities, but also
the radiationless (Auger) transition probabilities, to obtain the
quantities $W^{X}$.
 
A complete calculation of radiative and radiationless decay rates from
the double-K hole ground configuration was performed for Al and Sc.
Radiative transitions include K$\alpha^{\rm h}$ and K$\beta^{\rm h}$
hypersatellite lines, as well as K$\alpha\alpha$ (one electron - two
photon transitions) lines. The results are presented in Table
~\ref{tab_01b} together with the total radiative and radiationless
transition probabilities, for each of the two initial levels.

In Table ~\ref{tab_02} we provide the results of the complete
calculation of $W^X$ for the different lines in Al and Sc and of the
statistical average transition probability of line $X$, defined as the
quantity
\begin{equation} 
W^{X}_{SA}=\frac{1}{g\left(\gamma\right)}
\sum_i g\left(  i\right) W_{i}^{X}
\label{12}%
\end{equation} 

We observe that the values of $W^{X}$ and $W^{X}_{SA}$ are nearly
identical.  This results from the fact that Eq. (\ref{12}) can be
obtained from Eq.  (\ref{10}) if the summation $W_i +A_i $ has the
same value for all initial levels. This is the case of Al and Sc
presented in Table ~\ref{tab_01b}.

Total radiationless level widths are the sums of a large number of
transition rates. We may assume that the relativistic effects tend to
average out to some extent, similarly to what happens with total
radiative widths ~\cite{book_cra_1985}.  To test this assumption,
the value of radiationless (Auger) decay rate for each initial level
of the Ti 2s$^2$ 2p$^6$ 3s$^2$ 3p$^6$ 3d$^2$ 4s$^2$ ground
configuration was computed by adding the values of radiationless
transition probabilities for all levels of the final 1s 2p$^6$ 3s$^2$
3p$^6$ 3d$^2$ 4s$^2$ configuration. As shown in Table ~\ref{tab_03a},
no significant variation of the decay rates was found for different
initial levels, which shows that the radiationless decay rates do not
depend significantly on the particular level of the initial
configuration.
This validates the use of Eq.  (\ref{12}) as a good value to
$W^{X}$.

The ratio of the intensities of K$\beta_{1}^{\rm h}$ to K$\beta
_{3}^{\rm h}$ hypersatellite lines computed in this work, with and
without the Breit interaction, using the MCDF code of Desclaux and
Indelicato, are presented in Table ~\ref{tab_03}, together with the
values obtained by Chen \etal, which included Breit and
vacuum-polarization corrections. We notice that our values for these
ratios are larger than Chen's results for $Z\leq40$.  The two
approaches yield ratios that are in good agreement for $Z>40$.
 
In order to compare the transition energy values obtained in this work with 
experiment and calculations from other authors, we used the statistical 
average energy $E^{X}_{\rm SA}$ for the $X$ line defined in our previous 
article ~\cite{1105} as 
\begin{equation} 
E^{X}_{\rm SA}=\frac{1}{g\left(  \gamma\right)  }
\sum\limits_i g\left(  i\right)
\left( 
\frac{\sum\limits_{f^{X}}E_{if}W_{if}}{\sum\limits_{f^{X}}W_{if}}
\right)
.
\label{eq003} 
\end{equation} 
In this calculation we assumed that all $i$ levels of the $\gamma$
configuration are statistically populated. The quantity in parenthesis
is the average energy of the $X$ line, defined as the sum of the
energies of all individual $i\to f$ transitions in the $X$
line from an initial level $i$, $E_{if}$, weighted by the
corresponding $W_{if}$ \ radiative transition probability.
 
 
\subsection{Widths of 1s$^{-2}$ levels of Al and Sc} 
 
Using the values presented in Table ~\ref{tab_01b}, we were able to
calculate the widths of the 1s$^{-2}$ levels of Al and Sc, which are
displayed in Table ~\ref{tab_04}.

We believe this is the first time that level widths are calculated for
a double-K hole level. These values can be compared with existing
single-K hole level widths, using the expression 
$\Gamma \left(\rm{1s} ^{-2}\right) = 2\Gamma  \left(
\rm{1s} ^{-1}\right) $.
 
For Al, using the Evaluated Atomic Data Library (EADL) value
$\Gamma\left( \rm{1s} ^{-1}\right) =0.37$ eV proposed by Campbell and
Papp ~\cite{1219}, we obtain $\Gamma\left( \rm{1s} ^{-2}\right) =0.74$
eV, lower than the value calculated in this work. We note, however,
that the experimental values of $\Gamma\left( \rm{1s} ^{-1}\right) $
for Al referred to by the same authors are higher than their proposed
value.
Three of these values were derived from indirect measurements, using
other level widths to obtain the 1s$^{-1}$ level width. The only
experiment that led directly to the 1s$^{-1}$ level width yielded the
value 0.47 eV.

For Sc, the same authors propose $\Gamma\left( \rm{1s} ^{-1}\right)
=0.83$ eV, which yields $\Gamma\left( \rm{1s} ^{-2}\right) =1.66$ eV,
in excellent agreement with the value obtained in this work.
 
\subsection{Energy shifts} 
 
In Table ~\ref{tab_05} we present the results obtained in this work
for the K$\beta _{1}^{\rm{h}}$ and K$\beta_{3}^{\rm{h}}$ energy shifts
of the elements where we can distinguish these two lines.
 
This Table shows that our results for the K$\beta^{\rm{h}}$
hypersatellites energy shifts, relative to the corresponding diagram
line energies, are in good agreement with the results of Chen \etal
~\cite{308}, ours being smaller by less than 0.2 \% throughout.

To compare to the available experimental results, we present in Table
~\ref{tab_06} the K$\beta_{1,3}^{\rm{h}}$ energy shifts calculated in
this work.
Our results agree in general with experiment, as it can be seen in
Fig.~\ref{figure3}, although the uncertainties of the latter are very
large, with the exception of the recent experimental value of Diamant
\etal ~\cite{1015}.
 
\subsection{Breit interaction and QED corrections} 
 
To assess the contribution of the Breit interaction to the
K$\beta^{\rm{h}}$ hypersatellites energy shifts, we computed these
shifts with and without inclusion of the Breit interaction in the
calculation. We computed separately the K$\beta_{1}^{\rm{h}}$ energy
shifts, first with the Breit term (cf.  Eq. (\ref{eq:breit})) included
in the self-consistent process and the higher-order terms as a
perturbation after the self-consistent process is finished, and then
with Breit interaction neglected. The results are presented in Table
~\ref{tab_07}. Although Chen \etal~\cite{308} present their results
for these shifts, obtained with the Dirac-Slater approach, in graphic
form only, we easily conclude that our results, using the MCDF
approach, are in very good agreement with the results of Chen \etal.

The $\rm{K}\beta_{1}^{\rm{h}} $ to $ \rm{K}\beta_{3}^{\rm{h}}$
intensity ratio is sensitive to the inclusion of the Breit
interaction, similarly to Chem \etal~\cite{308} finding for the
$\rm{K}\alpha_{1}^{\rm{h}} $ to $\rm{K}\alpha_{2}^{\rm{h}}$ intensity
ratio. The inclusion of this interaction decreases the
K$\beta^{\rm{h}}$ intensity ratio at low $Z$ (21\% for $Z=18$) and
increases it for medium and high $Z$ ($\sim$ 5\% at $Z\simeq 50$).
However, since relativity affects the 3p$_{1/2}$ and 3p$_{3/2}$ levels
in a similar way the K$\beta^{\rm{h}}$ intensity ratio increases
monotonically towards the $jj$ coupling limit of 2.


The evolution of the relative contribution of Breit interaction to the
K$\beta_{1}^{\rm h}/$K$\beta_{3}^{\rm h}$ relative intensity ratio is
illustrated in Fig.~\ref{figure2}.

On the other hand, QED contributions for the energy shifts and
transition probabilities have been found to be negligible. For
instance, QED contributions for the $\rm{K}\beta_{1}^{\rm{h}}-
\rm{K}\beta_{1}$ energy shift in Hg is only 0.3\% compared with 13\%
from the Breit interaction contribution, whereas for the
$\rm{K}\beta_{1}^{\rm{h}}/\rm{K}\beta_{3}^{\rm{h}}$ intensity ratio
the QED contribution is 0.05\%, compared with 1.8\% from the Breit
contribution. The QED contributions for the $\rm{K}\beta_{1}^{\rm{h}}-
\rm{K}\beta_{1}$ energy shift are presented in Table~\ref{tab_07}.

 
\section{Conclusion} 
 
In this work we used the MCDF program of Desclaux and Indelicato to
compute energy shifts of the K$\beta_{1}^{\rm{h}}$ and
K$\beta_{3}^{\rm{h}}$ hypersatellite lines relative to the parent
diagram lines for several values of $Z$ throughout the periodic table.
One of the aims of this work was to assess the contribution of the
Breit interaction to these shifts. Our results confirm the earlier
findings of Chen \etal~\cite{308} for these shifts and extended them
to higher values of $Z$. We also calculated the $
\rm{K}\beta_{1}^{\rm{h} }$ to $ \rm{K}\beta_{3}^{\rm{h}}$ intensity
ratio for the same values of $Z$. Our results are significantly lower
than Chen \etal values for the same ratios, for $Z$ lower than $40$,
and agree with the values of these authors for higher values of $Z$.
The total widths of the double-hole K levels of Al and Sc were also
computed and our values were found in good agreement with the ones
obtained from proposed values of single-K hole levels ~\cite{1219}.

 
\ack 

This research was partially supported by the FCT project
POCTI/FAT/44279/2002 financed by the European Community Fund FEDER,
and by the TMR Network Eurotraps Contract Number ERBFMRXCT970144.
Laboratoire Kastler Brossel is Unit{\'e} Mixte de Recherche du CNRS
n$^{\circ}$ C8552.


\section*{References}


\begin{thebibliography}{99}

\bibitem{1105} Martins M C, Costa A M, Santos J P, Parente F and
  Indelicato P 2004 \jpb {\bf 37} 3785

\bibitem{888} Briand J P, Billy L, Charles P, Essabaa S, Briand P,
   Geller R, Desclaux J P, Bliman S and C. Ristori 1990 \PRL {\bf 65} 159 

\bibitem{889} Moribayashi K, Sasaki A and Tajima T 1998 \PR {\it A}
  {\bf 58} 2007

\bibitem{911} Briand J P, Chevallier P, Tavernier M and Rozet J P
  1971 \PRL {\bf 27} 777

\bibitem{900} Briand J P, Chevallier P, Johnson A,
   Rozet J P, Tavernier M and A. Touati A 1974 \PL {\it A} {\bf 49} 51

\bibitem{899} Briand J P, Touati A, Frilley M, Chevallier P, Johnson
A, Rozet J P, Tavernier M, Shafroth S and Krause M O 1976 \jpb  {\bf 9} 1055

\bibitem{1015}  Diamant R, Huotari S, H{\"a}m{\"a}l{\"a}inen K, Sharon R, Kao C C
and Deutsch 2003 \PRL {\bf 91} 193001

\bibitem{rep2000} Deutsch M, Huotari S, Hamalainen K, Diamant R and
  Kao C -C 2000 {\it ESRF Annual Report 2000} Experiment number HE-790
  
\bibitem{1208} Huotari S, H{\"a}m{\"a}l{\"a}inen K, Diamant R, Sharon
  R, Kao C C and Deutsch 2004 {\it J. Electron. Spectrosc. Related
    Phenomena} {\bf 137} 293

\bibitem{308} Chen M H, Crasemann B and Mark H 1982 \PR {\it A} {\bf
25} 391

\bibitem{32}  Desclaux J P 1975 {\it Comp. Phys. Commun.} {\bf 9} 31

\bibitem{62} Indelicato P  1996 \PRL {\bf 77} 3323

\bibitem{62a} Indelicato P and Desclaux J P, 2005 {\it MCDFGME, a
    MultiConfiguration Dirac Fock and General Matrix Elements program
    (release 2005)} http://dirac.spectro.jussieu.fr/mcdf

\bibitem{47} Indelicato P 1995 \PR {\it A} {\bf 51} 1132 

\bibitem{231} Ravenhall D E and Brow G E 1951 \PRS {\bf 208} 552

\bibitem{229} Sucher J 1980 \PR {\it A} {\bf 22} 348

\bibitem{230} Mittleman M H  1981 \PR {\it A} {\bf 24} 1167

1\bibitem{12} Akhiezer A I and Berestetskii V B 1965 {\it Quantum
    Quantum Electrodynamics} 

\bibitem{846} Lindgren I,  Persson H, Salomonson S and Labzowsky L 1995 \PR {\it A} {\bf 51} 1167

\bibitem{115} Mohr P J 1982 \PR {\it A} {\bf 26} 2338

\bibitem{114} Mohr P J and Kim Y -K 1992 \PR {\it A} {\bf 45} 2727

\bibitem{116} Mohr P J 1992 \PR {\it A} {\bf 46} 4421

\bibitem{58} Indelicato P, Gorceix O and Desclaux J P 1987 \jpb {\bf 20} 651

\bibitem{56} Indelicato P and Desclaux J P 1990 \PR {\it A} {\bf 42} 5139

\bibitem{53} Indelicato P, Boucard S and Lindroth E 1992 \PR {\it A}
   {\bf 46} 2426 

\bibitem{847} Indelicato P, Boucard S and Lindroth E 1998 {\it
   Eur. Phys. J. D} {\bf 3} 29

\bibitem{242} Indelicato P and Mohr P J 1991 {\it Theor. Chim. Acta} {\bf 80} 207

\bibitem{263} Blundell S A 1992 \PR {\it A} {\bf 46} 3762

\bibitem{288} Blundell S A 1993 \PR {\bf T46} 144

\bibitem{iam2001} Indelicato P and Mohr P J 2001 {\it Phys. Rev. A} {\bf 63} 052507

\bibitem{914} Boucard S and Indelicato P 2000 {\it Eur. Phys. J. D} {\bf 8} 59

\bibitem{158} Klarsfeld S 1969 \PL {\bf 30A} 382

\bibitem{277} Fullerton L W and G. A. Rinker Jr G A 1976 \PR {\it A} {\bf 13} 1283

\bibitem{book_cra_1985} B. Crasemann (ed) 1985 {\it Atomic Inner-Shell
    Physics} (New York, Plenum Press) p. 72

\bibitem{1219} Campbell J L and Papp T 2001 {\it At. Data Nucl. Data
    Tables} {\bf 77} 1


\end{thebibliography}


%

 
\newpage



\newpage

\Table{\label{tab_01b}Radiative transition probabilities of K$\alpha
  \alpha$, K$\alpha^{\rm h}$ and K$\beta^{\rm h}$ lines, Auger and
  total transition probabilities for each initial level ($LSJ_i$) in
  Al ($Z=13$) and Sc ($Z=21$). Numbers in parenthesis indicate a power
  of ten.}
\br
&  &\centre{5}{$W_i^X$ (s$^{-1}$)} & $W_i$ (s$^{-1}$) & $A_i$ (s$^{-1}$)\\
\ns
&&\crule{5}\\
& $LSJ_i$ & K$\alpha_2^{\rm h}$ & K$\alpha_1^{\rm h}$ & K$\beta_{13}^{\rm
  h}$ &
K$\alpha_2 \alpha_3$ & K$\alpha_1 \alpha_3$ \\
\mr Al & $^2$P$_{1/2}$ & $7.35(13)$ & $6.74(11)$ & $1.20(12)$ &
$9.37(10)$ & $5.45(7)$ &
$7.55(13)$ &  $1.56(15)$ \\
& $^2$P$_{3/2}$ & $6.36(13)$ & $5.40(11)$ & $1.20(12)$ & $1.00(11)$ &
$5.06(7)$ &
$6.55(13)$ &  $1.49(15)$ \\
Sc & $^2$D$_{3/2}$ & $4.93(14)$ & $2.98(13)$ & $7.04(13)$ & $4.01(11)$
& $1.69(9)$ &
$5.94(14)$ &  $1.91(15)$ \\
& $^2$D$_{5/2}$ & $4.88(14)$ & $2.18(13)$ & $7.01(13)$ & $3.92(11)$ &
$1.19(9)$ &
$5.81(14)$ &  $1.95(15)$ \\
\br
\end{tabular}
\end{indented}
\end{table}
 

\newpage

\Table{\label{tab_02}Comparison between the results of a complete
  calculation $W^X$ and a statistical average calculation $W_{\rm
    SA}^X$ of the transition probability for K$\alpha \alpha$,
  K$\alpha^{\rm h}$ and K$\beta^{\rm h}$ lines in Al ($Z=13$) and Sc
  ($Z=21$). Numbers in parenthesis indicate a power of ten.}
\br
&  & K$\alpha_2^{\rm h}$ & K$\alpha_1^{\rm h}$ & K$\beta_{13}^{\rm h}$ &  K$\alpha_2 \alpha_3$ & K$\alpha_1 \alpha_3$ \\
\mr
Al & $W^X$ (s$^{-1}$)  & $6.68(13)$ & $5.83(11)$ &  $1.20(12)$ &  $9.82(10)$ & $5.19(7)$  \\
& $W_{\rm SA}^X$ (s$^{-1}$) & $6.69(13)$ & $5.85(11)$ &  $1.20(12)$ &  $9.81(10)$ & $5.19(7)$  \\
Sc & $W^X$ (s$^{-1}$) & $3.96(14)$ & $2.50(13)$ &  $7.02(13)$ &  $3.96(11)$ & $1.39(9)$  \\
& $W_{\rm SA}^X$ (s$^{-1}$) & $3.96(14)$ & $2.50(13)$ &  $7.02(13)$ &  $3.95(11)$ & $1.39(9)$  \\
\br
\end{tabular}
\end{indented}
\end{table}


\newpage

\Table{\label{tab_03a} Auger decay rates (ADR) per initial level
  ($LSJ_i$) of the Ti ground configuration 2s$^2$ 2p$^6$ 3s$^2$ 3p$^6$
  3d$^2$ 4s$^2$ obtained by adding the values of radiationless
  transition probabilities for all levels of the final configuration
  1s 2p$^6$ 3s$^2$ 3p$^6$ 3d$^2$ 4s$^2$. } \br
$LSJ_i$ & ADR (s$^{-1}$) \\
\mr
$^3$P$_{0}$ & $1.592(14)$ \\
$^1$S$_{0}$ & $1.592(14)$ \\
$^3$P$_{1}$ & $1.594(14)$ \\
$^3$F$_{2}$ & $1.594(14)$ \\
$^1$D$_{2}$ & $1.594(14)$ \\
$^3$P$_{2}$ & $1.593(14)$ \\
$^3$F$_{3}$ & $1.594(14)$ \\
$^3$F$_{4}$ & $1.594(14)$ \\
$^1$G$_{4}$ & $1.593(14)$ \\

\br
\end{tabular}
\end{indented}
\end{table}


\newpage

\Table{%
Ratio of the K$\beta_{1}^{\rm h}$ to K$\beta_{3}^{\rm h}$ hypersatellite lines intensities, K$\beta_1^{\rm h}/$K$\beta_3^{\rm h}$, %
computed in this work, with and without %
the Breit interaction included, and compared with Chen \etal\protect~\cite{308}.
\label{tab_03}}\br
&  \centre{2}{K$\beta_1^{\rm h}/$K$\beta_3^{\rm h}$} \\
\ns
&\crule{3}\\
& \centre{2}{This work}   \\
&\crule{2}\\
$Z$ & Without  Breit & With Breit & Chen  \\
\mr
18 & 0.020 & 0.017 & 0.0093 \\
20 & 0.040 & 0.035 & 0.022  \\
29 & 0.32  & 0.38  &        \\
30 & 0.44  & 0.45  & 0.389  \\
36 & 0.84  & 0.88  & 0.831  \\
38 & 0.96  & 1.00  &       \\
40 &       &       & 1.10  \\
45 &       &       & 1.34  \\
46 & 1.33  & 1.39  &       \\
47 &       &       & 1.42  \\
48 & 1.39  & 1.46  &       \\
49 &       &       & 1.48  \\
54 & 1.55  & 1.62  & 1.61  \\
56 & 1.59  & 1.66  &       \\
60 &       &       & 1.72  \\
65 &       &       & 1.75  \\
70 & 1.77  & 1.82  &       \\
80 & 1.84  & 1.87  &       \\
\br
\end{tabular}
\end{indented}
\end{table}


\newpage

\Table{\label{tab_04} Values of widths for the 1s$^{-2}$ levels of Al
  and Sc calculated in this work. }
\br
& $LSJ_i$ & $\Gamma _i$ (eV) \\
\mr
Al & $^2$P$_{1/2}$ & $1.079$ \\ 
   & $^2$P$_{3/2}$ & $1.022$ \\ 
Sc & $^2$D$_{3/2}$ & $1.645$ \\ 
   & $^2$D$_{5/2}$ & $1.669$ \\ 
\br
\end{tabular}
\end{indented}
\end{table}


\newpage

\Table{%
K$\beta_{1}^{\rm{h}}$ and K$\beta_{3}^{\rm{h}}$%
  energy shifts, in eV, computed in this work and compared with Chen \etal\protect~\cite{308}.
\label{tab_05}}\br
&  \centre{2}{$E$(K$\beta_{1}^{\rm h}$)-$E$(K$\beta_{1}$) } & &
  \centre{2}{$E$(K$\beta_{3}^{\rm h}$)-$E$(K$\beta_{3}$)} & &
  $E$(K$\beta_{1}^{\rm h}$)-$E$(K$\beta_{3}^{\rm h}$) \\

\ns
&\crule{2} & &\crule{2} & &\crule{1} \\
 $Z$ & This work & Chen  & &  This work & Chen  & &  This work  \\
\mr
18 &  226.04 &  226.3  &    &   225.34 &  225.4  &  &   0.88 \\ 
20 &  255.22 &  255.5  &    &   254.37 &  254.3  &  &   1.24 \\ 
25 &         &  324.8  &    &          &  324.5  &  &        \\ 
29 &  381.62 &         &    &   381.14 &         &  &   2.96 \\ 
30 &  396.03 &  397.4  &    &   396.52 &  397.0  &  &   3.43 \\ 
31 &  412.10 &         &    &   412.15 &         &  &   3.65 \\ 
36 &  491.47 &  492.3  &    &   491.22 &  492.1  &  &   8.25 \\ 
38 &  524.43 &         &    &   524.17 &         &  &  10.87 \\ 
40 &         &  558.8  &    &          &  558.6  &  &        \\ 
45 &         &  645.7  &    &          &  645.2  &  &        \\ 
46 &  662.45 &         &    &   661.86 &         &  &  28.67 \\ 
47 &         &  681.8  &    &          &  681.2  &  &        \\ 
48 &  698.88 &         &    &   698.11 &         &  &  35.50 \\ 
49 &  717.42 &  718.8  &    &   716.56 &  718.0  &  &  39.22 \\ 
54 &  813.56 &  815.5  &    &   812.14 &  814.0  &  &  63.60 \\ 
56 &  854.28 &         &    &   852.22 &         &  &  76.43 \\ 
60 &         &  941.7  &    &          &  938.7  &  &        \\ 
65 &         & 1055.6  &    &          & 1051.1  &  &        \\ 
70 & 1175.10 &         &    &  1168.26 &         &  & 229.67 \\ 
80 & 1451.49 &         &    &  1438.14 &         &  & 444.43 \\ 
\br
\end{tabular}
\end{indented}
\end{table}


\newpage

\Table{\label{tab_06} The K$\beta_{1,3}^{\rm{h}}$ energy shifts
  calculated in this work and the available experimental results. }
\br
&  \centre{2}{K$\beta_{1,3}^{\rm h}-$K$\beta_{1,3}$ energy shifts (eV)} \\
\ns
&\crule{2}\\
 $Z$ & This work & Experiment  \\
\mr
13 & 157.82  &   \\ 
14 & 171.34  &   \\ 
15 & 184.40  &   \\ 
16 & 197.63  &   \\ 
17 & 211.39  &   \\ 
18 & 225.24  &   \\ 
20 & 254.16  &   \\ 
21 & 267.66  &   \\ 
23 & 295.36  &   \\ 
25 &   & $345 \pm 35^{\rm a}$  \\ 
26 & 337.75  & $336.0 \pm 0.5^{\rm b}$  \\ 
27 & 351.85  &   \\ 
28 & 366.46  &   \\ 
29 & 380.36  &   \\ 
30 & 395.58  &   \\ 
31 & 411.01  &  $390 \pm 20^{\rm c}$  \\ 
36 & 489.78  &   \\ 
38 & 522.60  &   \\ 
46 & 660.00  &   \\ 
48 & 696.26  &   \\ 
49 & 714.25  & $830\pm 60^{\rm c}$  \\ 
54 & 810.42  &   \\ 
56 & 850.82  &   \\ 
70 & 1169.63  &   \\ 
80 & 1443.46  &   \\ 
\br
$^{\rm a}$ Ref. ~\cite{900} \\
$^{\rm b}$ Ref. ~\cite{1015}\\ 
$^{\rm c}$ Ref. ~\cite{899} 
\end{tabular}
\end{indented}
\end{table}


\newpage

\Table{\label{tab_07}Breit and QED contributions to K$\beta_1^{\rm h}$
  energy shift.}  \br
     & \centre{2}{Breit} & & \centre{2}{QED} \\
     & \crule{2} & & \crule{2} \\
 $Z$ & (eV) & Percentage & & (eV) & Percentage  \\
\mr
18 & 1.66   &  0.7\%  & &   -0.08   &    -0.04\%  \\ 
20 & 2.31   &  0.9\%  & &   -0.10   &    -0.04\%  \\ 
29 & 8.59   &  2.0\%  & &   -0.30   &    -0.08\%  \\      
30 & 8.30   &  2.1\%  & &   -0.29   &    -0.07\%  \\ 
36 & 14.77  &  3.0\%  & &   -0.47   &    -0.09\%  \\ 
38 & 17.50  &  3.3\%  & &   -0.54   &    -0.10\%  \\ 
46 & 31.96  &  4.8\%  & &   -0.88   &    -0.13\%  \\ 
48 & 36.54  &  5.2\%  & &   -0.99   &    -0.14\%  \\ 
54 & 52.79  &  6.5\%  & &   -1.61   &    -0.20\%  \\ 
56 & 59.51  &  7.0\%  & &   -1.53   &    -0.18\%  \\ 
70 & 121.72 & 10.4\%  & &   -2.96   &    -0.25\%  \\ 
80 & 188.49 & 13.0\%  & &   -4.59   &    -0.32\%  \\ 
\br
\end{tabular}
\end{indented}
\end{table}
 
%
%
\Figures

\includegraphics{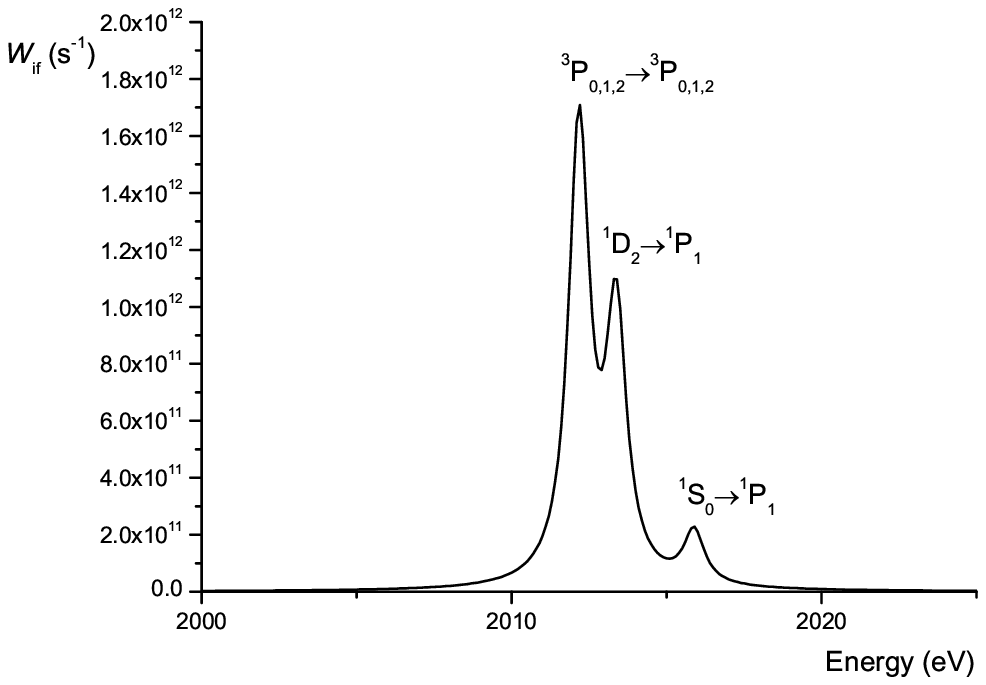}

\Figure{\label{figure} Calculated spectrum of Si K$\beta_{13}^{\rm h}$
  lines. A lorentzian with $\Gamma$=0.8 eV was used for each of the 14
  transitions, which yield this profile. }
%
%

\includegraphics{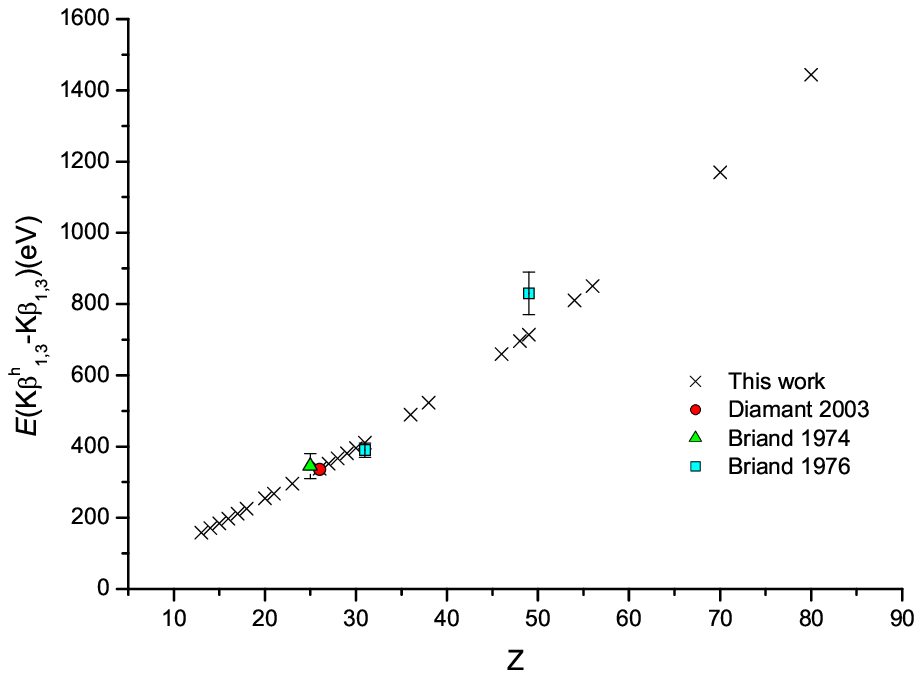}

\Figure{\label{figure3} Calculated K$\beta_{1,3}^{\rm{h}}$ energy
  shifts compared with available experimental results.}
%
%

\includegraphics{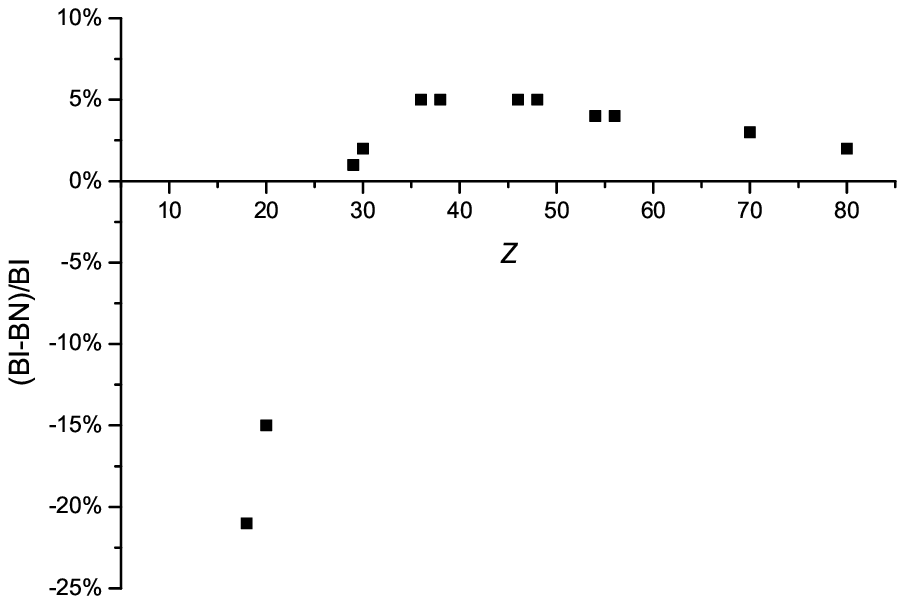}

\Figure{\label{figure2} Relative contribution of Breit interaction on
  the K$\beta_{1}^{\rm h}/$K$\beta_{3}^{\rm h}$ intensity ratio. BI
  and BN stand for Breit included and Breit neglected, respectively.}
%

\end{document}